\documentclass[prl,superscriptaddress, floatfix, twocolumn, showpacs,preprintnumbers,amsmath,amssymb]{revtex4}

\usepackage{epsfig}
\usepackage{dcolumn}
\usepackage{bm}

\newcommand {\BR}{\ensuremath{{\cal B}}}
\newcommand {\PR}{\ensuremath{PR}}

\newcommand {\Bp}{B^+}
\newcommand {\Bz}{B^0}
\newcommand {\Bs}{B_s^0}
\newcommand {\Bsb}{\overline{B}\vphantom{B}_s^0}
\newcommand {\Bzb}{\overline{B}\vphantom{B}^0}
\newcommand {\Dz}{D^0}
\newcommand {\Dm}{D^-}

\newcommand {\Dsm}{D_s^-}

\newcommand {\Dzb}{\ensuremath{\overline{D}\protect\vphantom{D}^0}}
\newcommand {\pip}{\pi^+}
\newcommand {\pim}{\pi^-}
\newcommand {\Kp}{K^+}
\newcommand {\Km}{K^-}
\newcommand {\BpToDPi}{\Bp \to \Dzb \pip}
\newcommand {\BzToDPi}{\Bz \to \Dm \pip}
\newcommand {\BsToDPi}{\Bs \to \Dsm \pip}

\newcommand {\um}{~\mu{\rm m}}
\newcommand {\GeVc} {~{\rm GeV}\!/c}
\newcommand {\MeVc} {~{\rm MeV}\!/c}

\newcommand {\SVXII}{SVX II}

\begin{document}

\title{ \boldmath Measurement of the Ratios of Branching Fractions 
$\BR(\BsToDPi)/\BR(\BzToDPi)$ and 
$\BR(\BpToDPi)/\BR(\BzToDPi)$ \unboldmath}
 
\affiliation{Institute of Physics, Academia Sinica, Taipei, Taiwan 11529, Republic of China} 
\affiliation{Argonne National Laboratory, Argonne, Illinois 60439} 
\affiliation{Institut de Fisica d'Altes Energies, Universitat Autonoma de Barcelona, E-08193, Bellaterra (Barcelona), Spain} 
\affiliation{Baylor University, Waco, Texas  76798} 
\affiliation{Istituto Nazionale di Fisica Nucleare, University of Bologna, I-40127 Bologna, Italy} 
\affiliation{Brandeis University, Waltham, Massachusetts 02254} 
\affiliation{University of California, Davis, Davis, California  95616} 
\affiliation{University of California, Los Angeles, Los Angeles, California  90024} 
\affiliation{University of California, San Diego, La Jolla, California  92093} 
\affiliation{University of California, Santa Barbara, Santa Barbara, California 93106} 
\affiliation{Instituto de Fisica de Cantabria, CSIC-University of Cantabria, 39005 Santander, Spain} 
\affiliation{Carnegie Mellon University, Pittsburgh, PA  15213} 
\affiliation{Enrico Fermi Institute, University of Chicago, Chicago, Illinois 60637} 
\affiliation{Joint Institute for Nuclear Research, RU-141980 Dubna, Russia} 
\affiliation{Duke University, Durham, North Carolina  27708} 
\affiliation{Fermi National Accelerator Laboratory, Batavia, Illinois 60510} 
\affiliation{University of Florida, Gainesville, Florida  32611} 
\affiliation{Laboratori Nazionali di Frascati, Istituto Nazionale di Fisica Nucleare, I-00044 Frascati, Italy} 
\affiliation{University of Geneva, CH-1211 Geneva 4, Switzerland} 
\affiliation{Glasgow University, Glasgow G12 8QQ, United Kingdom} 
\affiliation{Harvard University, Cambridge, Massachusetts 02138} 
\affiliation{Division of High Energy Physics, Department of Physics, University of Helsinki and Helsinki Institute of Physics, FIN-00014, Helsinki, Finland} 
\affiliation{University of Illinois, Urbana, Illinois 61801} 
\affiliation{The Johns Hopkins University, Baltimore, Maryland 21218} 
\affiliation{Institut f\"{u}r Experimentelle Kernphysik, Universit\"{a}t Karlsruhe, 76128 Karlsruhe, Germany} 
\affiliation{High Energy Accelerator Research Organization (KEK), Tsukuba, Ibaraki 305, Japan} 
\affiliation{Center for High Energy Physics: Kyungpook National University, Taegu 702-701; Seoul National University, Seoul 151-742; and SungKyunKwan University, Suwon 440-746; Korea} 
\affiliation{Ernest Orlando Lawrence Berkeley National Laboratory, Berkeley, California 94720} 
\affiliation{University of Liverpool, Liverpool L69 7ZE, United Kingdom} 
\affiliation{University College London, London WC1E 6BT, United Kingdom} 
\affiliation{Massachusetts Institute of Technology, Cambridge, Massachusetts  02139} 
\affiliation{Institute of Particle Physics: McGill University, Montr\'{e}al, Canada H3A~2T8; and University of Toronto, Toronto, Canada M5S~1A7} 
\affiliation{University of Michigan, Ann Arbor, Michigan 48109} 
\affiliation{Michigan State University, East Lansing, Michigan  48824} 
\affiliation{Institution for Theoretical and Experimental Physics, ITEP, Moscow 117259, Russia} 
\affiliation{University of New Mexico, Albuquerque, New Mexico 87131} 
\affiliation{Northwestern University, Evanston, Illinois  60208} 
\affiliation{The Ohio State University, Columbus, Ohio  43210} 
\affiliation{Okayama University, Okayama 700-8530, Japan} 
\affiliation{Osaka City University, Osaka 588, Japan} 
\affiliation{University of Oxford, Oxford OX1 3RH, United Kingdom} 
\affiliation{University of Padova, Istituto Nazionale di Fisica Nucleare, Sezione di Padova-Trento, I-35131 Padova, Italy} 
\affiliation{University of Pennsylvania, Philadelphia, Pennsylvania 19104} 
\affiliation{Istituto Nazionale di Fisica Nucleare Pisa, Universities of Pisa, Siena and Scuola Normale Superiore, I-56127 Pisa, Italy} 
\affiliation{University of Pittsburgh, Pittsburgh, Pennsylvania 15260} 
\affiliation{Purdue University, West Lafayette, Indiana 47907} 
\affiliation{University of Rochester, Rochester, New York 14627} 
\affiliation{The Rockefeller University, New York, New York 10021} 
\affiliation{Istituto Nazionale di Fisica Nucleare, Sezione di Roma 1, University di Roma ``La Sapienza," I-00185 Roma, Italy} 
\affiliation{Rutgers University, Piscataway, New Jersey 08855} 
\affiliation{Texas A\&M University, College Station, Texas 77843} 
\affiliation{Istituto Nazionale di Fisica Nucleare, University of Trieste/\ Udine, Italy} 
\affiliation{University of Tsukuba, Tsukuba, Ibaraki 305, Japan} 
\affiliation{Tufts University, Medford, Massachusetts 02155} 
\affiliation{Waseda University, Tokyo 169, Japan} 
\affiliation{Wayne State University, Detroit, Michigan  48201} 
\affiliation{University of Wisconsin, Madison, Wisconsin 53706} 
\affiliation{Yale University, New Haven, Connecticut 06520} 
\author{A.~Abulencia}
\affiliation{University of Illinois, Urbana, Illinois 61801}
\author{D.~Acosta}
\affiliation{University of Florida, Gainesville, Florida  32611}
\author{J.~Adelman}
\affiliation{Enrico Fermi Institute, University of Chicago, Chicago, Illinois 60637}
\author{T.~Affolder}
\affiliation{University of California, Santa Barbara, Santa Barbara, California 93106}
\author{T.~Akimoto}
\affiliation{University of Tsukuba, Tsukuba, Ibaraki 305, Japan}
\author{M.G.~Albrow}
\affiliation{Fermi National Accelerator Laboratory, Batavia, Illinois 60510}
\author{D.~Ambrose}
\affiliation{Fermi National Accelerator Laboratory, Batavia, Illinois 60510}
\author{S.~Amerio}
\affiliation{University of Padova, Istituto Nazionale di Fisica Nucleare, Sezione di Padova-Trento, I-35131 Padova, Italy}
\author{D.~Amidei}
\affiliation{University of Michigan, Ann Arbor, Michigan 48109}
\author{A.~Anastassov}
\affiliation{Rutgers University, Piscataway, New Jersey 08855}
% ...
\collaboration{CDF Collaboration}
\noaffiliation


\begin{abstract} 

We report an observation of the decay $\Bs \to \Dsm \pip$ in $p
\overline{p}$ collisions at $\sqrt{s} = 1.96~\rm{TeV}$ using 115~$\rm
pb^{-1}$ of data collected by the CDF II detector at the Fermilab
Tevatron. We observe $83 \pm 11$ $\Bs \to \Dsm \pip$ candidates, representing
a large increase in statistics over previous measurements
and the first observation of this decay at a $p \overline p$
collider. We present the first measurement of the relative branching
fraction
$\BR(\BsToDPi) / \BR(\BzToDPi)= \rm 1.32 \pm 0.18(\rm{stat.}) \pm 0.38(\rm{syst.})$.
We also measure  
$ \BR(\BpToDPi)/\BR(\BzToDPi) = \rm 1.97 \pm 0.10(\rm{stat.}) \pm
0.21(\rm{syst.})$,
which is consistent with previous measurements. 
\end{abstract}


\maketitle

\normalsize

$\Bs$-$\Bsb$ oscillation is expected to occur in the Standard Model of
particle physics. Measurement of the oscillation frequency, when
combined with that for $\Bz$-$\Bzb$ oscillation, tests the unitarity
of the Cabibbo-Kobayashi-Maskawa (CKM) quark mixing matrix. A deviation
from unitarity could arise from a variety of new physics effects
\cite{NewPhysicsMix}. $\Bs$ meson oscillations have yet to be directly
observed, with a lower limit (95\%~C.L.)  on the oscillation frequency
currently at  14.5~$\rm  ps^{-1}$~\cite{PDG}.  For comparison,   the
average $B$ meson  lifetime  is of the order   of 1 ps, and  the $B^0$
oscillation  frequency is 0.5~$\rm ps^{-1}$.  This lower limit implies
that excellent proper  time  resolution  is  required to  observe  the
oscillation. In semileptonic $\Bs$ decays, the proper time measurement
is  degraded   due to  the   undetected  neutrino. Fully reconstructed
hadronic  $\Bs$ decays  such as  $\BsToDPi$  do not  suffer  from this
problem~\cite{BWorkshop}.  Obtaining a large  sample of such decays is
an important first step toward measuring $\Bs$ mixing.

\begin{figure}
\epsfig{file=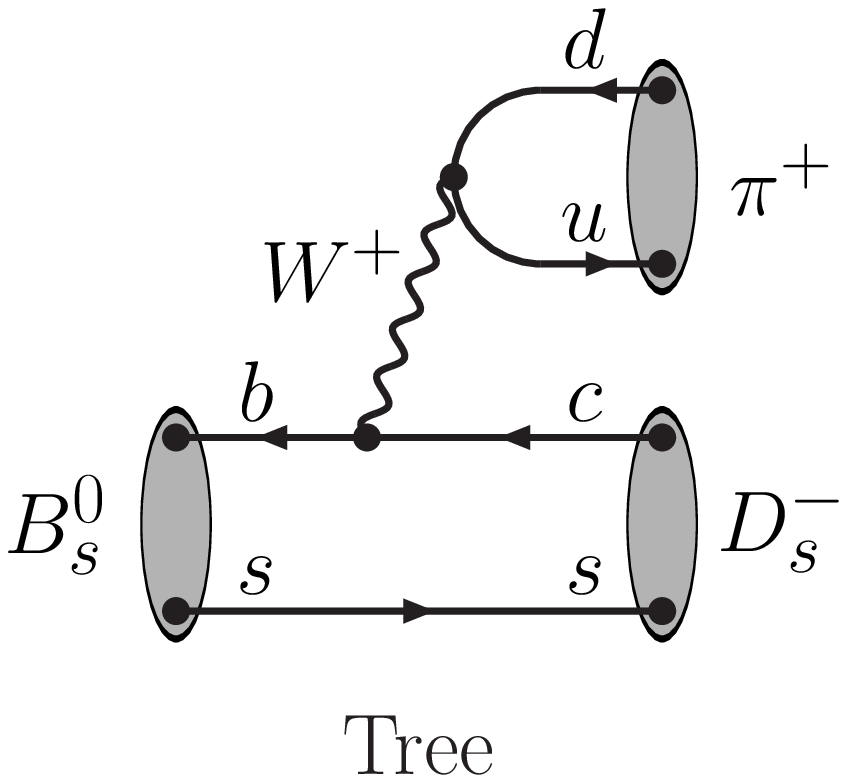,     width = 0.33 \linewidth}\nolinebreak
\epsfig{file=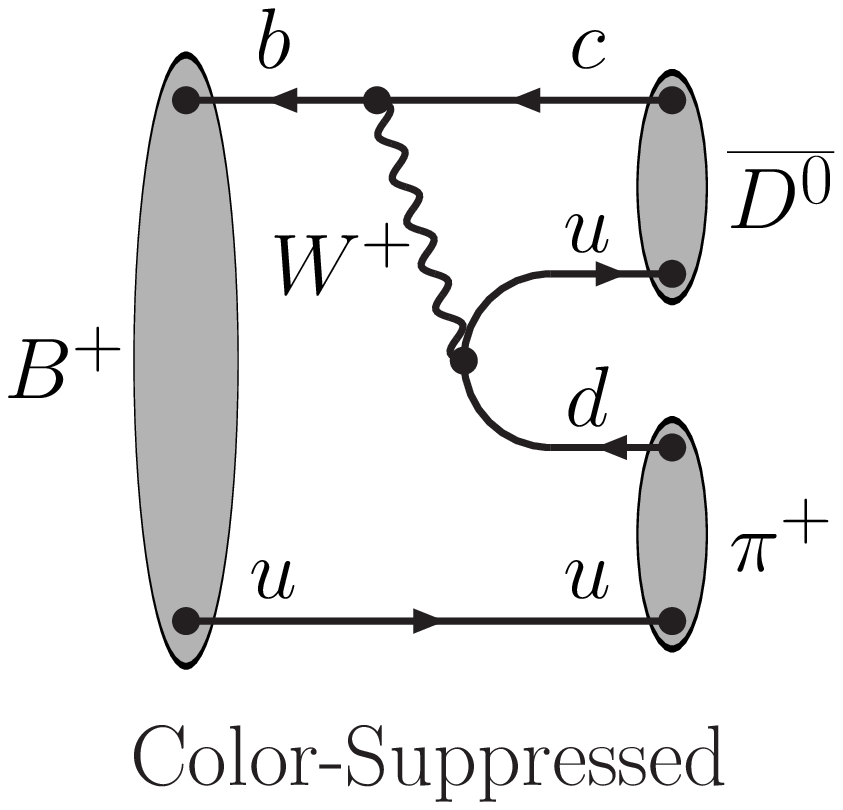,    width = 0.33 \linewidth}\nolinebreak
\epsfig{file=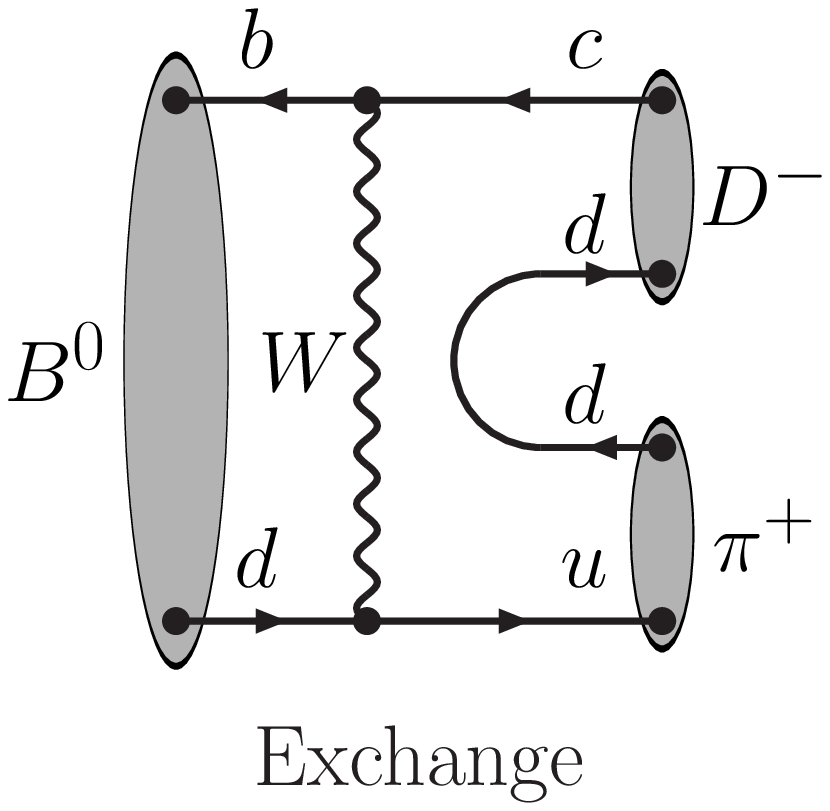, width = 0.33 \linewidth}
\caption{Different Feynman diagrams which contribute to $B \to D \pi$ decays.
\label{fig:diagrams}}
\end{figure}

In addition, the measurement of the branching ratio of $\BsToDPi$
decay along with that of $\BzToDPi$ and $\BpToDPi$ can be used to
study $B$ meson decay mechanisms. As shown in
Fig.~\ref{fig:diagrams}, $\BsToDPi$ decay proceeds only at tree
level while the $\Bz$ and $\Bp$ modes have additional, non-tree,
contributions. Therefore, measurements of ratios of branching
fractions of these decays can, in principle, isolate contributions
from the different decay diagrams~\cite{IainStewart}.

To date, a few $\BsToDPi$ events have been observed~\cite{Aleph,Opal}
in $e^+e^-$ collisions at LEP. $B$ factories, currently running at the
$\Upsilon(4S)$ resonance, do not produce $\Bs$ mesons. However, large
samples of $\Bs$ are produced at the Tevatron. The ability to trigger
on displaced vertices allows the upgraded Collider Detector at
Fermilab (CDF II) to collect large samples of fully reconstructed
$\Bs$ decay modes.

In this letter, we present an observation of $\BsToDPi$ decays and a
measurement of the ratios of branching fractions of $\BsToDPi$ and
$\BpToDPi$ relative to the branching fraction of $\BzToDPi$ decays. We
use a sample of fully reconstructed $B \to D \pi$ decays corresponding
to 115 pb$^{-1}$ of $p \overline p$ collisions at $\sqrt{s} = 1.96
~\rm TeV$ collected by the CDF II detector at the Fermilab Tevatron
between February 2002 and January 2003. Charge conjugate modes are
implied throughout this paper.

The components of the CDF II detector relevant to this analysis are
described briefly below; a more complete description can be found
elsewhere \cite{inclBSpectrum}. We use tracks reconstructed by both
the Central Outer Tracker (COT) and the silicon microstrip detector
(\SVXII) in the range $|\eta| \leq 1$ \cite{Tracking}, where $\eta$ is
the pseudorapidity defined as $\eta = - \ln \tan (\theta/2)$ and
$\theta$ is the polar angle with respect to the proton beam
direction. The {\SVXII} detector consists of double-sided silicon
strip sensors arranged in five cylindrical shells with radii between
2.5 and 10.6 cm
\cite{SVXII}. Surrounding the {\SVXII} is the COT \cite{COT}, an
open-cell drift chamber that has an inner (outer) radius of 40 (137)
cm. The COT has 96 layers, organized in 8 superlayers with alternating
superlayers of axial and $\pm 2^\circ$ stereo readout.
The $B$ decays used in this analysis are selected with a three-level
trigger system. At Level 1, charged tracks are reconstructed in the
COT axial superlayers by a hardware processor, the eXtremely Fast
Tracker (XFT) \cite{XFT}. This trigger requires two oppositely charged
tracks with transverse momenta $p_T \geq 2\GeVc$ and scalar sum
$p_{T1} + p_{T2} \geq 5.5\GeVc$.  At Level 2, the Silicon Vertex
Trigger (SVT) \cite{SVT} associates {\SVXII} $r-\varphi$ position
measurements with XFT tracks. This provides a precise measurement of
the track impact parameter, $d_0$, which is defined as the projection
of a track's distance of closest approach to the beam line onto the
transverse plane. The resolution of the impact parameter measurement
is $50 \um$, which includes approximately $30 \um$ contribution from
the transverse beam size. Hadronic decays of heavy flavor particles
are selected by requiring two tracks with $120 \um
\leq d_0 \leq 1000\um$. The two trigger tracks must have an
opening angle in the transverse plane satisfying $2^\circ
\leq |\Delta \varphi| \leq 90^\circ$ and must satisfy the requirement
$L_{xy} > 200 \um$, where $L_{xy}$ is defined as the distance in the
transverse plane from the beam line to the two-track vertex projected onto the
two-track momentum vector. A complete event reconstruction is
performed at Level 3, and the Level 1 and Level 2 requirements are
confirmed. Since events satisfying the displaced vertex trigger are
dominated by the decays of promptly produced charm, the largest
background to $B \to D \pi$ decays results from combining a prompt $D$
meson with an unrelated track from the event.

Reconstruction of $B$ meson candidates begins by selecting $D$
meson candidates. We reconstruct the following $D$ meson decay modes:
$\Dzb\to \Kp\pim$, $\Dm\to \Kp\pim\pim$ and $\Dsm\to\phi\pi^-$ followed by 
$\phi \rightarrow K^+K^-$. We exploit the narrow $\phi\to K^+K^-$ resonance 
in the $\Dsm$ decays to greatly suppress background by requiring 
$1010 \MeVc ^2 < m(K^+ K^-) < 1028 \MeVc ^2$. No particle identification is 
used in this analysis. All particle hypotheses consistent with the 
candidate decay structure are attempted. The track combinations which comprise
a $D$ meson candidate are required to originate from a common vertex. An 
additional track with $p_T > 1.6 \GeVc$ and $\Delta R(D, \pi_B) < 1.5$ is 
added and assigned a pion mass to reconstruct the $B$ meson candidate. We 
define $\Delta R = \sqrt{\Delta \eta^2 + \Delta \varphi^2}$, where $\Delta
\varphi$ and $\Delta \eta$ are the azimuthal angle and pseudorapidity
of the pion with respect to the direction of the $D$ meson
candidate. The selection requirements are optimized to yield the
largest $S /\sqrt{S+B}$ for each of the decays. In the optimization
procedure, the number of signal events ($S$) is estimated using a
Monte Carlo simulation, while the number of background events ($B$) is
estimated using sidebands of the $B$ meson mass spectra reconstructed
in data.

\begin{figure}
\begin{minipage}{0.48 \linewidth}
\begin{center}
\epsfig{file = 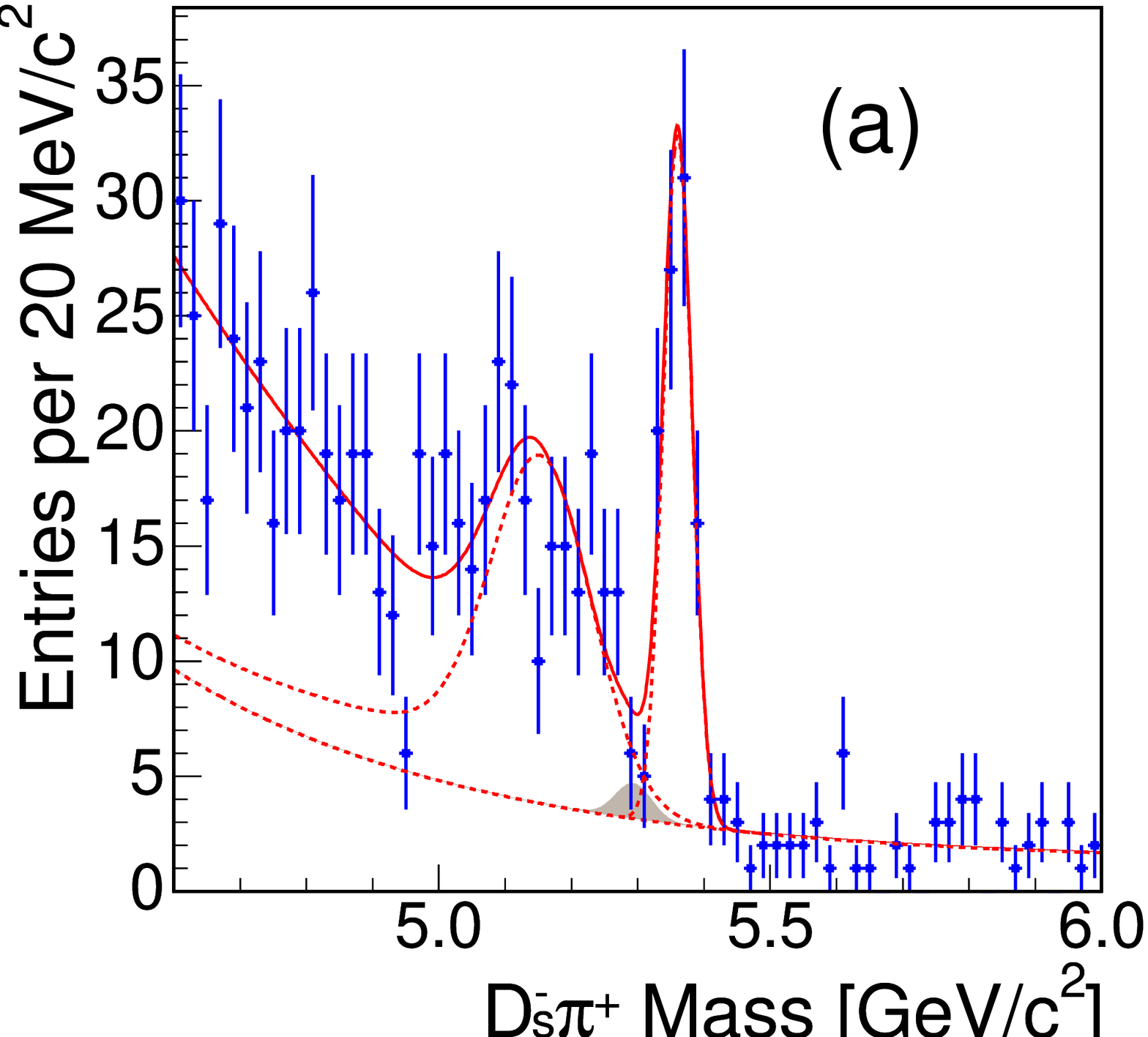, width = \linewidth}
\end{center}
\end{minipage}\nolinebreak
\begin{minipage}{0.48 \linewidth}
\begin{center}
\epsfig{file = 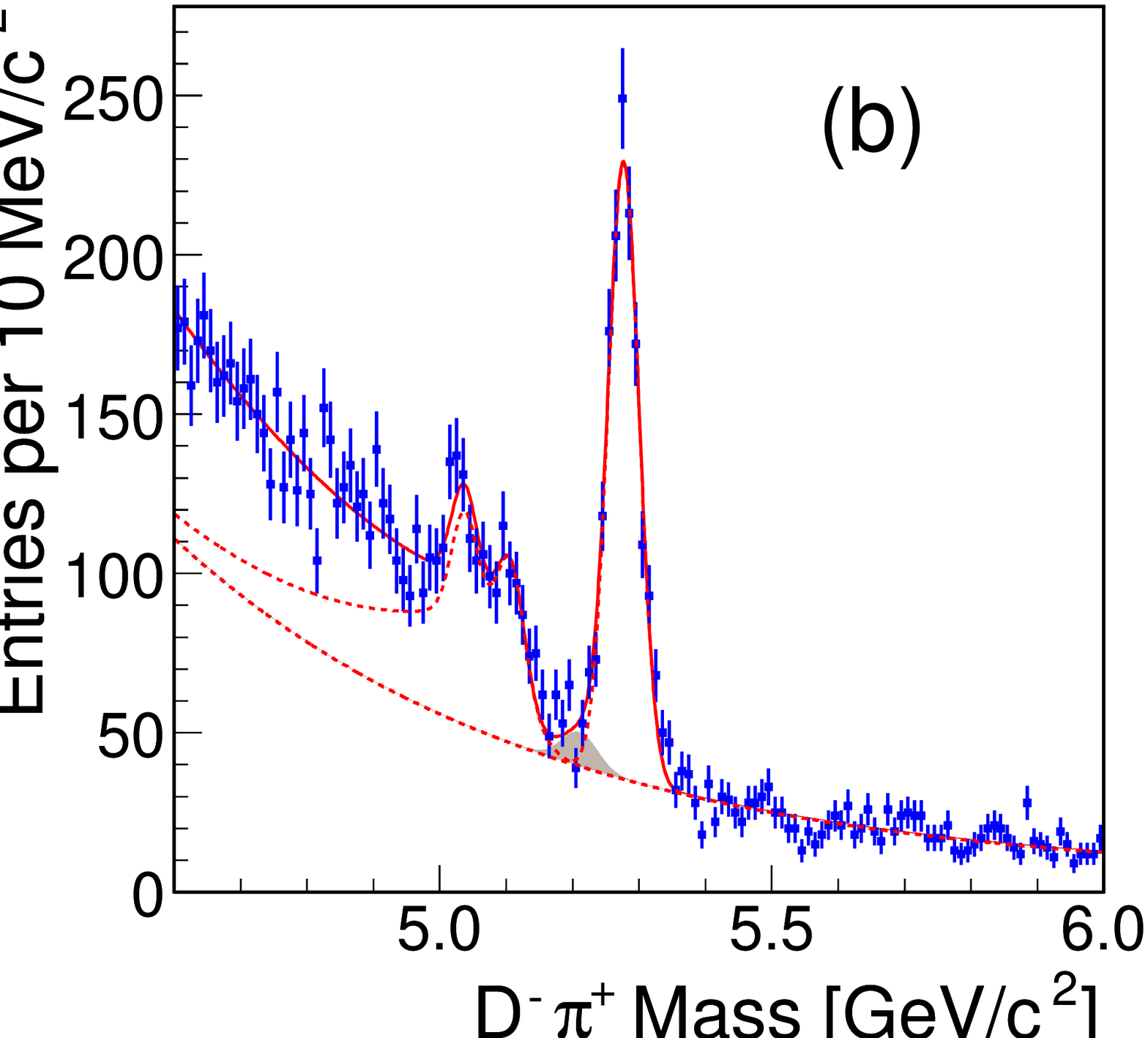, width = \linewidth}
\end{center}
\end{minipage}\\
\begin{minipage}{0.48 \linewidth}
\begin{center}
\epsfig{file = 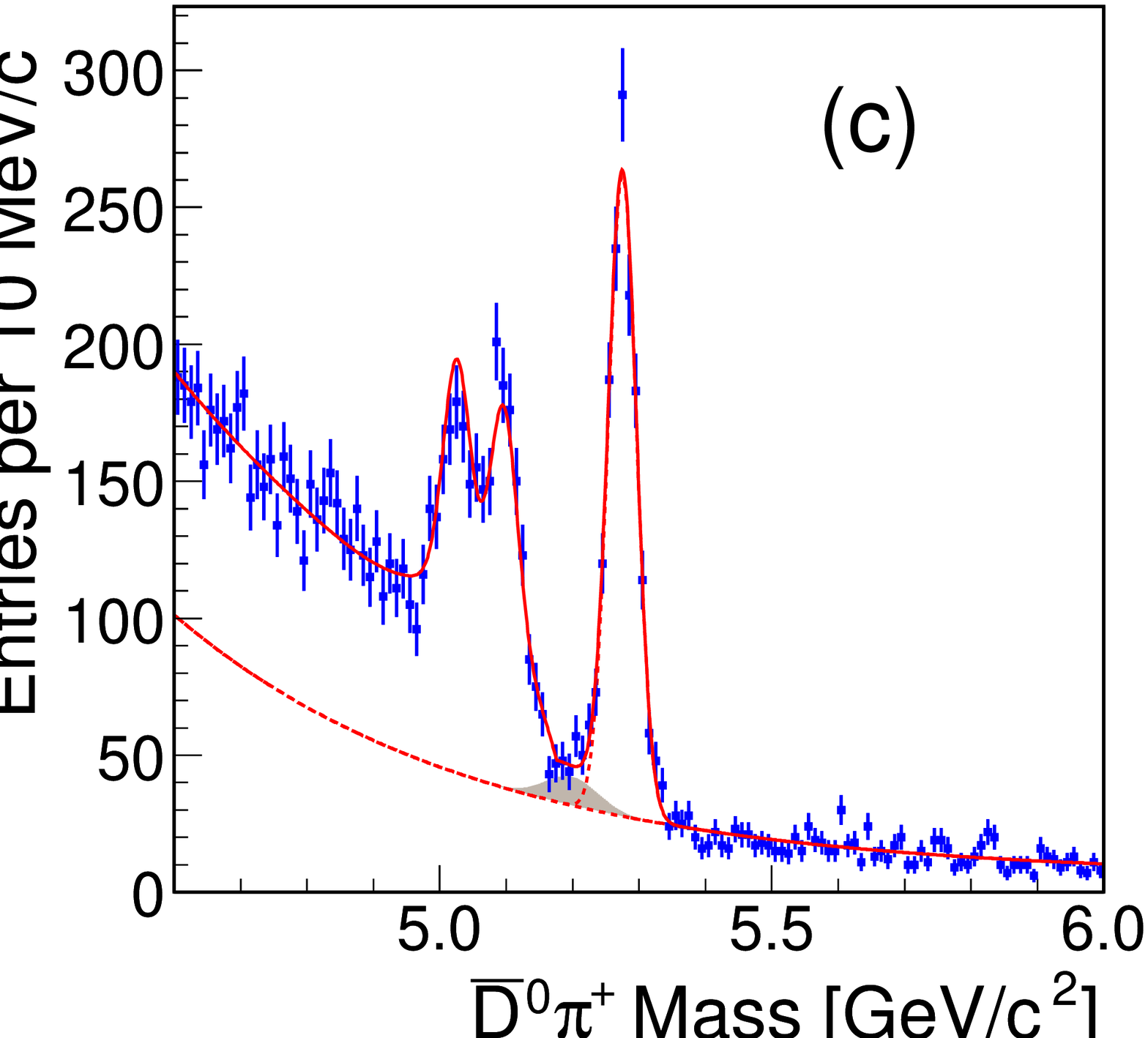, width = \linewidth}
\end{center}
\end{minipage}\nolinebreak
\begin{minipage}{0.48 \linewidth}
\begin{center}
\epsfig{file = 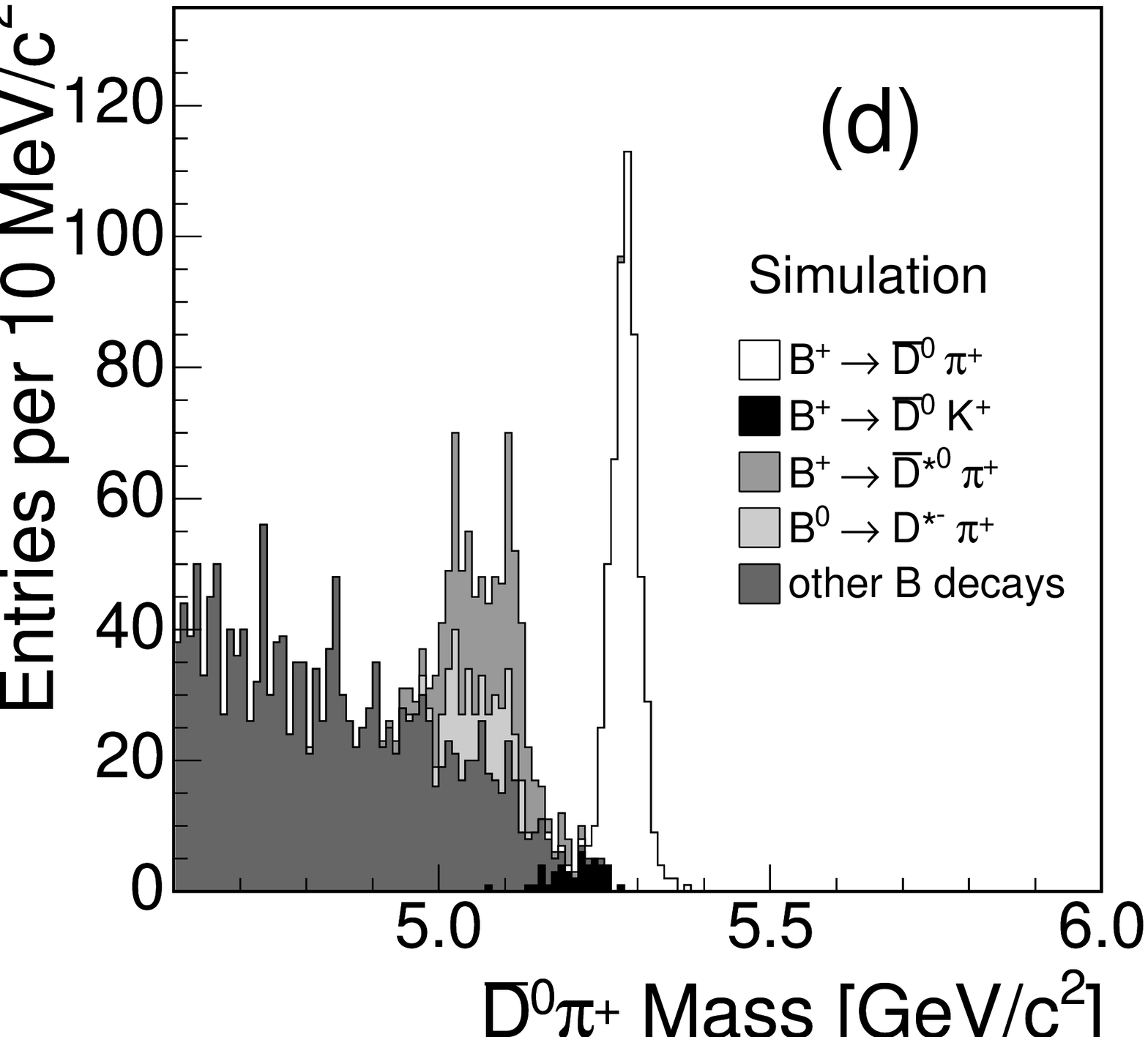, width = \linewidth}
\end{center}
\end{minipage}
\protect \caption{
Invariant mass spectra for (a) $\Dsm \pip$ (b) $\Dm \pip$ and (c)
$\Dzb \pip$ for data, with $\chi^2$ fits overlaid. The main background
under the signal peak is combinatorial, modeled with a combination of
a linear and exponential function. The shaded peak corresponds to $B
\to D K$ decays, and is modeled with a Gaussian. The additional
background component in the low mass region is due to partially
reconstructed $B$ decays. This background component is modeled using
shapes determined by inclusive $B \to D X$ Monte Carlo. A sample Monte
Carlo mass distribution for $\BpToDPi$ events is shown in
(d). Contributions from different sources in (d) are ordered
by peak position, from right to left.}
\label{fig:b-yields}
\end{figure}

We require the $B$ meson candidate tracks to be consistent with the
following constraints: the $D$ meson tracks originate from a common
vertex, the momentum of the $D$ meson points back, in three
dimensions, to the remaining $B$ meson candidate track, and the
invariant mass of the $D$ meson decay products is consistent with the
world average of the corresponding $D$ meson mass \cite{PDG}. We also
require that at least two of the $B$ meson daughter tracks are
consistent with the trigger requirements. The combinatorial background
is strongly reduced by requiring that $L_{xy}^B > 400 \um$ and
$L_{xy}(B \to D) > -150 \um$, where the latter refers to the $L_{xy}$
of the $D$ meson decay vertex with respect to the $B$ meson decay
vertex. The impact parameter of the $B$ meson momentum with respect to
the beam axis is required to be less than $80 \um$ to assure that the
$B$ meson candidate originates from the primary vertex. The invariant
mass distributions of $\Bs$, $\Bp$, and $\Bz$ meson candidates are
shown in Fig.~\ref{fig:b-yields}. The prominent peak at each
expected $B$ meson mass establishes $B \to D\pi$ decay signals,
including our observation of $\BsToDPi$.

We measure the following ratio:
\begin{widetext}
\begin{eqnarray}
\frac{\BR(\Bs \to \Dsm \pip)}{\BR(\Bz \to \Dm \pip)} = 
\frac{ N(\Bs)}{N(\Bz)} \cdot \frac {\epsilon(\Bz)}{\epsilon(\Bs)} \cdot 
\frac{f_d}{f_s} \nonumber \cdot 
\frac{ \BR ( \Dm \to \Kp \pim \pim ) }{ \BR ( \Dsm \to \phi \pim ) \cdot \BR(
	\phi \to \Kp \Km ) } \label{eqn:main}
\end{eqnarray}
\end{widetext}
where $N(\Bs)$ and $N(\Bz)$ are the signal yields, $f_d/f_s$ is the
ratio of fragmentation fractions for $\Bz$ and $\Bs$ mesons, and
$\epsilon(\Bz)/\epsilon(\Bs)$ is the ratio of trigger and
reconstruction efficiencies. Equation \ref{eqn:main} also applies to
the $\Bp \to \Dzb \pi^+$ mode, with terms for $\Bs$ decays replaced by
those relevant to $\Bp$ decays. There are three components to the
ratio of branching fractions measurement: the ratio of $B$ meson
yields obtained from fits to the invariant mass spectra, the ratio of
signal efficiencies obtained from Monte Carlo simulation, and the
product of previously measured production and branching fractions.

In our analysis, we use Monte Carlo simulation to determine shapes of
mass spectra and relative efficiencies. The Monte Carlo generation
proceeds as follows. Transverse momentum and rapidity distributions of
single $b$ quarks are generated based on calculations using NLO perturbative
QCD \cite{NDE}. $B$ meson kinematic distributions are obtained by
simulating Peterson fragmentation
\cite{Peterson} on quark-level distributions. Additional fragmentation 
particles, correlated $b-\overline{b}$ production and the underlying
event structure are not simulated. $B$ meson decays are simulated
using EvtGen~\cite{EvtGen}. The simulation of the CDF II detector and
trigger is based upon a GEANT description \cite{GEANT} that includes
effects of time variation of the beam position and hardware
configuration of the {\SVXII} and SVT.

The yield of $B$ mesons is extracted from the invariant mass spectra
using a binned $\chi^2$ fit, as shown in Fig.~\ref{fig:b-yields}. 
There are three contributions to the background
shape used in the fit. The combinatorial background component is
modeled with a linear combination of a linear and exponential
distribution. The shape of background due to Cabibbo-suppressed $B \to
D K $ decay is modeled by a Gaussian and is shown as shaded
distributions in Fig.~\ref{fig:b-yields}(a)-(c). Backgrounds due to
partially reconstructed $B$ decays contribute to the low mass side of
the signal peak and are modeled by inclusive $B \to D X$ Monte Carlo
simulation. The structures in the region close to the signal are due
to $B\to D^* \pi$ decays, where a photon or a $\pi^0$ from the $D^*$
decay is not reconstructed.  The decay of the polarized $D^*$ produces
the double-peaked structure. As an example of the various contributing
backgrounds, Fig.~\ref{fig:b-yields}(d) shows the invariant mass
spectrum for $B \to D X$ Monte Carlo reconstructed as $B^+ \to \Dzb
\pi^+$. In the fit to the data mass spectrum, the fractions of
combinatorial background and partially reconstructed $B$'s are allowed
to float in the fit. The ratio of Cabibbo-suppressed $B \to D K$
background to the corresponding signal is fixed to the world average
ratio of branching fractions, with the trigger and reconstruction
efficiencies of the two modes taken into account. The fitted yields of
$B^0 \to D^- \pi^+$, $B^+ \to \Dzb$ and $B_s^0 \to D_s^- \pi^+ $
decays are $1118\pm 43 ({\rm stat.})$, $1260 \pm 42 ({\rm stat.}) $
and $83\pm 11 ({\rm stat.})$ events, respectively.

Trigger and reconstruction efficiencies for $\Bz, \Bp$, and $\Bs$
decays are determined using Monte Carlo simulation. The trigger
efficiency differs among the three $B$ meson decay modes due to
differences in decay kinematics (e.g. opening angle distributions)
which arise from the different masses and spins of the intermediate and final
state particles. The ratios of efficiencies are $\epsilon(B^0) /
\epsilon(B^+) = 0.708 \pm 0.010 $ and $\epsilon(B^0) / \epsilon(B_s^0)
= 0.903\pm 0.012$, where the uncertainties are due to the limited statistics
of Monte Carlo samples.

World average values \cite{PDG} are used for the various branching
fractions in Equation \ref{eqn:main}. In terms of $B$ meson
production, we assume $f_d=f_u$, consistent with previous
measurement~\cite{fsfd}. The $\Bs/\Bz$ production fraction used in
this analysis is the world average value, $f_s/f_d = 0.270 \pm 0.029$,
currently dominated by LEP measurements. The ratio of fragmentation
fractions may be different in a hadron collider environment than in
$e^+e^-$ collisons. However, the ratio of fragmentation fractions
measured by CDF~\cite{fsfd} is consistent with the LEP results.

Many systematic uncertainties cancel in the measurement of ratios of
branching fractions due to the similarity of final state
kinematics. Systematic uncertainties come from three main sources:
fitting the invariant mass distributions to obtain signal yields,
determination of the ratio of efficiencies, and uncertainties on
external inputs. The contribution to the systematic uncertainty from
externally measured quantities is calculated by propagating world
average uncertainties. Systematic uncertainties on the signal yields
are determined by comparing the fitted yields after changing the
invariant mass region of the fit and varying the background shape
within the range allowed by the $B \to D X$ Monte Carlo statistics and
world average uncertainties on the branching fractions of
participating $B$ decays. Systematic uncertainties on the ratio of
efficiencies come from physics sources such as the choice of $p_T$
spectrum or meson lifetime, and detector sources such as inaccuracies
in the XFT and SVT hardware simulation. The uncertainty due to a given
source is estimated by the shift of the ratio of efficiencies when the
effect of that source is modified in the Monte Carlo simulation. The
effect of the choice of $B$ meson $p_T$ spectrum is estimated by
re-weighting the Monte Carlo to match the measured $B$ hadron $p_T$
spectrum \cite{inclBSpectrum}. Since the trigger and analysis
selection only accept events in which the $B$ meson is displaced from
the primary interaction point, the efficiencies depend upon the $B$
and $D$ meson lifetimes. To estimate uncertainties due to $B$ and $D$
meson lifetimes, the Monte Carlo is re-weighted with different
lifetimes within the world average uncertainties. Due to the different
specific ionization of $\pi^\pm$ and $K^\pm$ in the COT, kaons are 6\%
less efficient in satisfying the XFT requirements \cite{DtoHH}. The
Monte Carlo simulation is re-weighted to reproduce this effect. To
estimate the uncertainty due to imperfections in simulating the
signal selection requirement efficiencies, we compare efficiencies between
Monte Carlo simulation and sideband subtracted $\Bp$ and $\Bz$ signal
for every selection requirement separately.  In the case of the
$\Bp$/$\Bz$ branching fraction ratio measurement, there is an
additional systematic uncertainty associated with the fact that the
$\Bz$ decay has four tracks in the final state while $\Bp$ has only
three. We infer the corrections and systematic uncertainties due to
the fourth track by comparing trigger and reconstruction efficiencies
in data and Monte Carlo for semileptonic $B \to D^{*-}l^{+}X, D^{*-}
\to \Dzb \pim$ decays between the $\Dzb \to \Kp \pim$ and $\Kp \pim
\pip \pim$ final states.

The systematic uncertainties are summarized in Table~\ref{tab:system}. The 
total systematic uncertainty is obtained by adding individual contributions in
quadrature. Using Equation~\ref{eqn:main}, we obtain the following values for 
the ratios of branching fractions:
\begin{eqnarray}
\frac{\BR(B_s^0 \to D_s^- \pi^+)}{ \BR(B^0 \to D^- \pi^+)} = & 
1.32 \pm 0.18 (\rm{stat.}) \pm 0.10(\rm{syst.}) \nonumber \\  
& \pm 0.34 (\rm{BR}) \pm 0.14 (\rm{\PR})
\label{eqn:bsResult}
\end{eqnarray}
\begin{eqnarray}
\frac{\BR(B^+ \to \Dzb \pip)}{ \BR(B^0 \to D^- \pi^+)} = &
1.97 \pm 0.10 (\rm{stat.}) \pm 0.15(\rm{syst.}) \nonumber \\ 
&\pm 0.14 ({\rm BR})
\label{eqn:bpResult}
\end{eqnarray}
where $\rm BR$ and $\rm PR$ refer to the uncertainty on the ratio of $D$ meson
branching fractions and $\Bs$ meson production relative to $\Bz$,
respectively. Under the assumption of isospin invariance, our
measurement of the ratio $\BR(\BpToDPi)/\BR(\BzToDPi)$ is consistent
with the world average~\cite{PDG}.  This provides a high statistics
crosscheck of the measurement procedure.

\begin{table}
\caption{Summary of relative systematic uncertainties.
\label{tab:system}
}
\begin{ruledtabular}
\begin{tabular}{l|c|c}
Effect              & $B_s^0/B^0 [\%] $ & $B^+/B^0 [\%] $ \\
\hline
\hline
fit $N(B)$          & $  4.5 $ & $ 2.0 $ \\
fit $N(B^0)$        & $  2.3 $ & $ 2.3  $ \\
\hline
$B$ $p_T$ spectrum  & $ 2.6  $ & $  1.5  $ \\
trigger simulation  & $ 1.6$ & $ 0.2$ \\
selection simulation & $ 4.0  $ & $ 4.0  $ \\ 
$\phi^0$ mass cut   & $ 0.2$ &  not applicable \\
\hline
$\Bp/\Bs$ meson lifetime  & $ 2.0$ & $ 0.4$ \\ 
$\Dz/\Dsm$ meson lifetime  & $ 0.1$ & $ 0.1$ \\ 
\hline 	             
$\Bz$ meson lifetime & $ 0.4$ & $ 0.4$ \\ 
$\Dm$ meson lifetime & $<0.1 $   & $<0.1 $   \\ 
\hline	             
fourth track accept  &   not applicable  & $ 5.2  $ \\ 
\hline
Total               & $ 7.4  $ & $ 7.4  $ \\
\end{tabular}
\end{ruledtabular}
\end{table}

In conclusion, we have presented the first observation of $\BsToDPi$
decays in $p \overline{p}$ collisions, and the first measurement of
the $\BsToDPi$ branching fraction relative to the $\BzToDPi$ branching
fraction. The precision of this measurement is currently not adequate
to separate the contributions of different decay diagrams
\cite{IainStewart}. We expect the measurement precision to improve as
world average values of $D$ meson branching fractions and $B$ meson
production fractions improve. 

We thank the Fermilab staff and the technical staffs of the
participating institutions for their vital contributions. This work
was supported by the U.S. Department of Energy and National Science
Foundation; the Italian Istituto Nazionale di Fisica Nucleare; the
Ministry of Education, Culture, Sports, Science and Technology of
Japan; the Natural Sciences and Engineering Research Council of
Canada; the National Science Council of the Republic of China; the
Swiss National Science Foundation; the A.P. Sloan Foundation; the
Bundesministerium f\"ur Bildung und Forschung, Germany; the Korean
Science and Engineering Foundation and the Korean Research Foundation;
the Particle Physics and Astronomy Research Council and the Royal
Society, UK; the Russian Foundation for Basic Research; the Comisi\'on
Interministerial de Ciencia y Tecnolog\'{\i}a, Spain; in part by the
European Community's Human Potential Programme under contract
HPRN-CT-2002-00292; and the Academy of Finland.



\begin{thebibliography}{99}
\bibitem{Tracking}{CDF II uses a cylindrical coordinate system in which $\varphi$ is the
azimuthal angle, $r$ is the radius from the nominal beamline, $y$
points up and $z$ points in the proton beam direcion with the origin
at the center of the detector. The transverse plane is the plane
perpendicular to the $z$ axis. A superconducting magnet provides a
nearly uniform 1.4 T axial field in which charged particles are
reconstructed.}
\end{thebibliography}
\end{document}